# A quantitative analysis of the chain fountain

J. Pantaleone
Department of Physics
University of Alaska Anchorage

The chain fountain is an entertaining, counterintuitive phenomenon. When a chain flows up over the edge of a container and then falls to the ground below, it is observed that the top of the chain rises up above the container's edge. Here the steady-state motion of the fountain is analyzed experimentally and theoretically. Measurements are given for the speeds and heights for three different chains and three different distances from the container to the floor. It is shown theoretically that the distance the chain rises above the container is proportional to the force from the container on the chain. The basic physics behind this force is readily understandable and can be illustrated with simple examples. To quantitatively predict the chain's motion a simple model is developed for how the chain interacts with the container. This model shows that a link lifts-off from the container after rotating by a relatively small angle. The model's predictions agree very well with the measurements for the two ball-chains.

## I. INTRODUCTION

The motion of falling chains has been studied for a long time[1-20]. Chains are often used as an example of a continuous mass system. In addition, recent observations have found some very counterintuitive effects. For example, a chain striking a table may accelerate downward *faster* than a chain in free fall[11-13]. Also, when a pile of chain on a table is rapidly pulled horizontally, the chain may spontaneously rise up to form an arch above the table[14,15]. Most spectacular of all is the chain fountain[17-20], also known as "self-siphoning beads". In a popular video, Steven Mould[17] demonstrated that a chain falling from a container can spontaneously grow in height, and that the steady-state height can be far above the container, see Fig. 1. This spectacular, counterintuitive demonstration provides many educational opportunities.

The basic physical reason why the chain fountain occurs is straightforward[18]. When a link at rest in the container is pulled upward by the rising chain, this force accelerates the center of mass of the link upwards and also causes the link to rotate about the center of mass. The rotation of the link causes the end of the link in contact with the container floor (or chain pile) to push downwards, and so there is a corresponding upward "reaction" force on the link from the container. This reaction force is what causes the chain to rise up above the container.

Part of what makes the chain fountain such a good educational demonstration is that it can be explained using simpler physical examples. In particular, the interaction between the container and the link lifting off from it can be directly illustrated by a bullet-block experiment[23,24]. Consider what happens when an upward traveling bullet strikes a block resting on a horizontal surface. The block will rise *higher* when the bullet strikes the block near the end, than when the bullet strikes the block in the center[21-24]. The explanation here is identical to the chain fountain, the reaction force from the horizontal surface, produced by the block's rotation, pushes the block higher. A video of this bullet-block demonstration, including a discussion of how it relates to the chain fountain, is available online[24]. A discussion of the bullet-block experiment makes the chain fountain demonstration accessible to all levels of physics students.

There are only a few published papers which discuss the chain fountain. Initial work by Biggins and Warner[18] explained the basic reason for the chain rising above the container, measured how high it rose as a function of the distance from the container to the floor, and developed a crude model to describe the results. In a following work, Biggins described the shape of the chain fountain and the rate of approach to the steady state. More recently, Andrews et al.[20] have observed the chain fountain for very long chains and report a nonlinear relationship between the height of the fountain and the drop height.

In this paper the steady-state chain fountain is quantitatively analyzed. Section II contains our observations of chain fountains, with measured steady-state heights and also speeds for three different chains: two ball chains and one bar chain. Section III contains our theoretical model describing the chain fountain. Section III.A describes a general framework for the forces on the chain fountain. A formula is derived that connects the observed quantities---the chain speed and the chain rise and fall heights. Also, this section shows that the interaction between the container and the link lifting off from the container is crucial for the chain rising above the container. Then this interaction is examined in detail in Section III.B. Here formulas are derived for the height the chain rises to above the container, and for the critical link angle, in terms of the width and length of the chain links. The model's predictions agree well with the measured results for the two ball chains. For the bar chain the predictions are much less accurate, but the reason for the discrepancy is explained. Section IV discusses additional physical effects and also different ways the chain fountain can be used for teaching. Section V summarizes the results.

## II. OBSERVATIONS

A typical chain fountain is shown in Fig. 1. It consists of relatively straight, vertical sides with a small curved section at the top. It is common for waves to propagate along the chain (see right hand side of Fig. 1) and, near the top, small distortions from a smooth curve can be relatively long lasting. The height the chain rises to above the container fluctuates because of varying conditions in the container. For example, tangles or snags in the container temporarily reduce the height.

The chain fountain observations reported here were made using three different chains: a small-ball chain, a large-ball chain and a bar-chain, see Fig. 2. These chains were purchased from Ball Chain Manufacturing. The balls and bars are hollow and are connected by small rods that are flared on the ends. For all three chains, the maximum bending angle at a junction between a rod and a ball/bar was about 20 degrees. However, because of the connection geometry, the larger the bending angle the more the rod must slide inside the ball/bar. The bar chain and the large-ball chain were each about 30 meters long while the small-ball chain was 75 meters long. The physical parameters of these chains are given in Table 1 (see Fig. 3 for definitions of lengths).

To observe the chain fountain, a chain was placed in a clear, plastic container that had a base of 9 cm by 9 cm and a height of 13 cm, see Fig. 1. The chain was allowed to fall into a cardboard box that had a cloth towel on the bottom of it. For each chain type, the chain fountain was observed for three different average heights below the container ($h_{below}$): 1.29 m, 2.35 m and 3.57 meters and at least four runs were performed at each average fall height.

The rise of the chain above the container was video recorded and the video were analyzed using the freely available software Tracker[25]. From each video, measurements were made of the heights $h_{above}$ and $h_{below}$. $h_{above}$ is defined to be the steady state distance the chain rises above the top of the chain pile in the container and $h_{below}$ is defined to be the distance from the top of the chain pile to the floor. To qualify as being in a steady-state, it was required that the value of $h_{above}$ be the same for several frames. For each video, from one to five different set of heights were recorded. The smaller number of data points per video were recorded for the larger values of $h_{below}$ (because the chain was moving faster and so the experiment lasted for a shorter time) and for the small-ball chain (because it tended to tangle more often). From the values of $h_{above}$ and $h_{below}$, the dimensionless ratio $h_{above}/h_{below}$ was calculated. The average and standard deviation of this ratio is plotted in Fig. 4.

Also shown in Fig. 4 is the ratio of the steady-state chain speed to the free-fall speed. The free-fall speed was calculated using

$$V_{free-fall} = \sqrt{2gh_{below}} \tag{1}$$

This is the speed the chain would have if mechanical energy were conserved. To measure the steady-state chain speed, $V$, the cardboard box that the chain impacted was placed on a PASCO Force Platform. The force applied to the platform was recorded while the chain was falling on to it. The average rate of change of the force sensor was measured, $dW_{box}/dt$, during the steady state motion. Then the steady-state speed of the falling chain was calculated using

$$\frac{dW_{box}}{dt} = \lambda g V \tag{2}$$

where $V$ is the steady-state chain speed, $\lambda$ is the mass per unit length of the chain, and $g$ is the acceleration of gravity. The weight $W_{box}$ will generally have a dynamic contribution[3,11] proportional to $\lambda V^2$, however this will not contribute to the measurement in Eq. (2).

In general, the data show that the dimensionless ratios of $h_{above}/h_{below}$ (upper Fig. 4) and $V/V_{free-fall}$ (lower Fig. 4) are independent of the fall height, $h_{below}$. The average measured value of these dimensionless ratios, for each chain type, are given in Table 2. These results are consistent with those in Ref. 18, which measured a value $h_{above}/h_{below} = 0.14$ for a chain with dimensions similar to the large-ball chain used here, and with those is in Ref. 20 which found a similar value when $h_{below} < 3.5$ m.

## III. THEORETICAL EXPLANATIONS

### A. Forces on the chain fountain

A sketch of the steady-state forces acting on the chain fountain is shown in Fig. 5. For calculation purposes the chain has been broken up into the three sections: the part in the container, the part above the container, and the part below the container. The steady state forces acting on each section have been included on this sketch. The forces acting on the part of the chain in the container are shown on the left side of the sketch, those acting on the part above the container are shown in the middle and those acting on the part below the container are shown on the right side.

For the section below the container there are two forces included (shown on right side of the Fig. 5), the weight of this section of chain, $W_{below}$, and the tension in the chain at the top of this section, $F$. Any force acting on the bottom end of the chain, where it impacts the ground, has been neglected. This is because the chain will not support an upward (compressive) force, and any downward (tension) force, such has been observed in some experiments[11,12], is small and not crucial to understanding the chain fountain. Because this section of the chain is not accelerating, the forces must balance, so

$$F = W_{below} \qquad (3)$$

The other two sections of the chain are accelerating because although the speed is constant the direction of the velocity is not. To describe the chain's motion in these regions, a specialized form of Newton's law appropriate for a *fixed volume* of a continuous mass system will be used. Summing $\vec{f} = m\vec{a}$ over the chain in a fixed region, and assuming the system is in steady state, one gets

$$\vec{f}_{net} = Q\Delta \vec{v} \qquad (4)$$

Here $\vec{f}_{net}$ is the net force on the region, $Q$ is the mass flow per unit time through the boundaries, $\vec{v}$ is the velocity of the chain at the boundaries and $\Delta$ denotes the change between the boundaries. Eq. (4) is the steady state version of what is commonly known as the momentum principle or the momentum theorem. Because the system is in steady-state motion, the total momentum in a fixed region does not change. Then the net force acting on a region is equal to the flux of momentum out of the region. The momentum flux is the velocity, $\vec{v}$, times the mass flow per unit time through the boundary, $Q = \lambda v$, where $\lambda$ is the linear mass density of the chain. For a detailed discussion of Eq. (4) for a fluid, see e.g. Ref. 26.

For the section of chain above the container there are three forces acting on it (and shown in the middle of Fig. 5) the weight of the chain, $W_{above}$, and, at the bottom boundary, the two tension forces in the vertical sections of the chain. Since the chain is in steady-state motion these two tension forces must have the same value, $F$. Thus applying Eq. (4) to this region gives

$$2F + W_{above} = 2\lambda V^2 \qquad (5)$$

The left-hand side of Eq. (5) is the total downward force on the chain in this region and the right-hand side is the downward change in the momentum flux, $\lambda V^2$ going down minus the upward going $\lambda V^2$.

For the section of chain in the container, we focus on the few chain links in the transitional region where they change from horizontal and at rest to vertical and moving upwards. The force $F$ is pulling this section of the chain upwards at the top boundary of the region and the container supplies an additional upward force on the chain, $F`$, at the lower boundary of this region (and shown on the left side of Fig. 5). Any horizontal forces acting on this section of the chain have been neglected. The neglect of the horizontal forces here should be reasonable because the chain is relatively unconstrained in the horizontal direction, and because there is no net orientation of links in the chain pile so some of the effects of the horizontal forces should tend to cancel out. Applying Eq. (4) to this region gives

$$F + F' = \lambda V^2 \tag{6}$$

Here $F`$ is the average force acting upwards on the bottom of this region, F is the average force acting upwards on the top of the region, and $\lambda V^2$ is the momentum flux up and out of that region. The gravitational force acting on this region is neglected because this section only contains a few links and so their weight will be small compared to the other forces in Eq. (6).

The dimensionless ratio of the steady-state chain speed to the free fall speed can be calculated by combining Eqs. (1), (3) and (5) to give

$$\frac{V}{V_{\text{free-fall}}} = \sqrt{\frac{1}{2}\left(1 + \frac{W_{above}}{2W_{below}}\right)} \approx \sqrt{\frac{1}{2}\left(1 + \frac{h_{above}}{h_{below}}\right)} \tag{7}$$

The last part of Eq. (7) makes use of $W_{\text{below}} = \lambda h_{\text{below}} g$ and $W_{\text{above}} \approx \lambda 2 h_{\text{above}} g$. The last statement is a slight approximation because it neglects the curve in the top section. Note that Eq. (7) relates the two observations plotted in Fig. 4. Thus it gives a consistency check on the model. Also, the derivation of Eq. (7) did not include Eq. (6), so it is *independent of how the links interact in the container*.

The values of $V/V_{\text{free-fall}}$ predicted by Eq. (7), and as calculated from the observed values of $h_{\text{above}}/h_{\text{below}}$, are given in Table 2. The predicted values are slightly larger, but generally in reasonable agreement, with the measured values. Note that the small systematic discrepancy is too large to be explained by the approximation of neglecting the curvature in the chain at the top. Also, it must be pointed out that adding a downward force on the bottom of the chain on the right-hand side (as was observed in Refs. 11 and 12 for a chain in free-fall) would make this small discrepancy worse. The discrepancy could be due to a drag force acting on the moving chain.

The dimensionless ratio of the rise height to the fall height can be calculated by combining Eqs. (3), (5) and (6) to give

$$\frac{F'}{F} = \frac{W_{above}}{2W_{below}} \approx \frac{h_{above}}{h_{below}} \tag{8}$$

For the last part of Eq. (8), the deviations from vertical at the top of the chain fountain have again been neglected. Eq. (8) shows that the chain rises above the pile directly because of the reaction force, $F\grave{\ }$. To calculate $F\grave{\ }$, a model is needed for the interaction between the link and the container floor in order to determine how an upward pull of average size $F$ on one end of the link produces an upward force of average size $F\grave{\ }$ on the other end of the link.

Conservation of mechanical energy in falling chains has been a topic of much discussion. The current consensus in the literature is that mechanical energy should be conserved when chains unfold smoothly but, when chains move in or out of a pile, mechanical energy is not conserved[8,18]. For the chain fountain, since $h_{above}/h_{below} < 1$, Eq. (7) predicts that $V/V_{free-fall} < 1$, and thus the system does not conserve mechanical energy. The mechanical energy dissipation is occurring in the container, as the bottom link in the container is lifted off of the pile by the forces $F$ and $F$'. There the work done on that link increases both the rotational and translation kinetic energies of that link. However while the upward going chain has a net momentum it does not have a net angular momentum. Because the links in the container have no net orientation, for every link rotating clockwise there will be another link rotating counter-clockwise. Thus the rotational angular momentum and the rotational kinetic energy must be dissipated through inelastic interactions between the neighboring links. This interpretation agrees with the observation that, in the limit $F = F$', no rotation of the bottom link occurs and Eqs. (7) and (8) predict that mechanical energy is conserved.

As a check, note that Eq. (5) can be applied near the top of the chain to give $T \approx \lambda V^2$, where $T$ is the tension there. This result for a rotating chain or string can be found in several textbooks (see e.g. Ref. 27). It is relevant for discussions of the classroom demonstration of a spinning chain loop rolling along the floor[28].

**B. Detailed model of the link-container interaction**

The motion of the link in contact with the container floor (or pile) is determined by the forces and torques acting on it. Here a link is defined as two spheres connected by a rod, as shown in Fig. 3. Applying Newton's laws to the center of mass motion of the link gives

$$f + f' = M\frac{dv}{dt} \tag{9}$$

where $M$ is the mass of the link and lower case symbols are now used to represent the time-dependent versions of the analogous quantities in the previous section. The forces of the previous section are time averaged values: $F = \langle f \rangle$ and $F\grave{\ }=\langle f\grave{\ }\rangle$, and the steady-state chain velocity, $V$, is the final velocity of the link's center of mass, $V = v_{final}$. For the rotation of the link about its center of mass the equation of motion is

$$\left[f\left(\frac{L+D}{2}\right) - f'\left(\frac{L}{2}\right)\right]\cos\theta = I_{cm}\frac{d\omega}{dt} \tag{10}$$

Here $\theta$ is the angle between the link and the horizontal, $\omega = d\theta/dt$ is the angular velocity of the link about the center of mass, and $I_{cm}$ is the rotational inertia about the center of mass. Fig. 3 shows the definitions of the distances $L$ and $D$, and also where the force $f$ and $f\grave{}$ are located. The force $f$ is assumed to act at the outer middle part of the hemispherical section, and thus the perpendicular lever arm is the distance $((L+D)/2)\cos\theta$, while the force $f\grave{}$ acts at the lowest point of the hemisphere, where it is in contact with a horizontal surface. Since the contact point is always directly below the center of the hemisphere, the perpendicular lever arm about the center of mass is $(L/2)\cos\theta$.

It should be noted that Eq. (10) assumes a link to be a rigid structure, so its application to a ball chain is an approximation. Also note that for a ball chain the two spheres that constitute a link changes during the lift-off process. As a sphere leaves the horizontal surface, that sphere and the sphere below it that is still in contact with the surface, constitute one link. After a short time the lower sphere of this link lifts off the surface, at which point it becomes the upper sphere of the next link.

The linear and rotational motions of the link, as described by Eqs. (9) and (10), are not independent because the link is in contact with the container floor (or pile). Thus the container floor constrains the vertical velocity of the link there to be zero

$$v_{y,\text{bottom}} = v - \omega \frac{L}{2}\cos\theta = 0 \tag{11}$$

Note that here the lever arm is the same as in Eq. (10) for the lower side. Combining equations (9), (10) and (11), and solving for $f\grave{}$ at an arbitrary angle $\theta$ and angular velocity $\omega$, gives

$$f' = \left[\frac{1+(D/L)-\delta}{1+\delta}\right]f - \frac{[2+(D/L)]\delta\sin^2\theta}{[\cos^2\theta+\delta][1+\delta]}f - \frac{\delta ML\omega^2\sin\theta}{2[\cos^2\theta+\delta]} \tag{12}$$

Here the dimensionless quantity $\delta$ is defined as

$$\delta = \frac{4I_{cm}}{ML^2} \tag{13}$$

which is a measure of the rotational inertia about the center of mass. The first term in Eq. (12) is the initial reaction force when $\theta = 0$, the other terms account for the decrease in this quantity as the link rotates upward. This decrease occurs because the lever arm decreases and because the angular acceleration increases.

The dimensionless parameter $\delta$ depends on the shape of the link. For the link configuration shown in Fig. 3, and neglecting the mass of the connecting rods compared to the mass of the spheres, the parameter $\delta$ is approximately

$$\delta \approx 1 + \frac{2}{3}\left(\frac{D}{L}\right)^2 \tag{14}$$

For the bar chain, a link is chosen to be the simplest possible generalization of Fig. 3, where each ball is simply replaced by a bar. Each bar is assumed to be two hemispheres connected by a cylinder. Then the parameter $\delta$ is given by

$$\delta \approx \left\{\left(1-\frac{S}{L}\right)^2 + \frac{M_S}{M}\left[\frac{2}{3}\left(\frac{D}{L}\right)^2 + \left(\frac{S}{L}\right)\left(\frac{D}{L}\right) + \left(\frac{S}{L}\right)^2\right] + \frac{M_C}{M}\left[\frac{1}{2}\left(\frac{D}{L}\right)^2 + \frac{1}{3}\left(\frac{S}{L}\right)^2\right]\right\} \quad (15)$$

Here $M_S$ is the total mass in the (hemi-) spheres, $M_C$ is the total mass in the cylindrical sections, $M = M_S + M_C$, and the mass in the connecting rod has again been neglected. Taking the density and the thickness of the metal in the links to be the same everywhere, then these quantities cancel in the mass ratios so that one can take $M_S \propto \pi D^2$ and $M_C \propto \pi DS$. For the chains used here the values of $\delta$ given by Eqs. (14) and (15) are listed in Table 1.

Initially, the link is horizontal ($\theta = 0$) and at rest ($\omega = 0$) and the second and third terms in Eq. (12) vanish. Then $f\,\grave{}$ is a small, positive number, as expected from observations. As the link rotates upward, $\theta$ and $\omega$ both increase, and the second and third terms in Eq. (12) both cause $f\,\grave{}$ to decrease. Before $\theta$ reaches $\pi/2$ there is a critical angle, $\theta_c$, where $f\,\grave{} = 0$. For times (angles) after this critical point, the link will lift off the container floor and the constraint equation on the bottom end of the link, Eq. (11), is no longer valid, so Eq. (12) is also no longer valid. When the link lifts off of the container floor, then it becomes the piece of the chain supplying the upward force $f$ to the next link which has one end on the container floor.

To find the ratio of the rise height to the fall height as given in Eq. (8), the average of $f\,\grave{}$ is needed over the time the link is in contact with the floor. This average would be difficult to determine exactly, but it is not difficult to calculate approximately. The approximation is made that the contribution of the second and third terms in Eq. (12) are small compared to the first term, and that they can be treated as perturbations. This approximation is reasonable since the bar is moving slowest at the initial time where $\theta = 0$ and so the first term will contribute most to the time average.

Neglecting the second and third terms in Eq. (12), and assuming $f$ is a constant in time (so $f = F$), then Eqs. (9-11) can be solved exactly to give

$$\omega^2 \approx \left[\frac{4F}{ML}\right]\frac{[2+(D/L)]}{[1+\delta]}\sin\theta \quad (16)$$

The assumption that $f$ is a constant may seem drastic, however the time dependence of $f$ only affects the perturbative correction term. Thus this time dependence can be considered a higher order correction that is negligible at the level of the first correction term. Substituting this expression for $\omega^2$ into Eq. (12) allows the second and third terms to be combined to give

$$\frac{f'}{F} \approx \frac{1+(D/L)-\delta}{1+\delta} - \frac{3[2+(D/L)]\delta\sin^2\theta}{[\cos^2\theta+\delta][1+\delta]} \quad (17)$$

This equation can now be easily solved for the critical angle by setting $f' = 0$, to obtain

$$\sin^2\theta_c \approx \frac{[1+(D/L)-\delta][1+\delta]}{[1+(D/L)+\delta\{5+3(D/L)\}]} \tag{18}$$

The critical angle (the angle the link makes with the container floor when it starts to lift off from the floor) is an observable quantity. Note that the critical angle is independent of the fall height, $h_{below}$, and $\lambda$, and just depends on how the mass is distributed in the links of the chain through the parameters $D/L$ and $\delta$.

The values of the critical angle for the chains used here, as calculated using Eq. (18), are given in Table 2. These values are rather small. This has two important implications. First, note that these angles are smaller than the amount of bending possible at the junction between a ball and a connecting bar. Thus this justifies the choice of a link as consisting of the smallest possible unit, two balls. Second, it means that the small angle approximation can be used when calculating the time average of the last term in Eq. (17).

In the small angle approximation $\cos^2\theta \approx 1$ in the denominator of Eq. (17), $\sin^2\theta \approx \theta^2$ in the numerator, and the angular acceleration is constant over the interval. Thus $\theta^2$ depends on time to the fourth power, so the time average of $\theta^2$ reduces to $\theta_c^2/5$. Then the ratio of the heights above and below the container, as given by Eq. (8), is

$$\frac{h_{above}}{h_{below}} \approx \frac{[1+(D/L)-\delta]}{[1+\delta]}\left\{1-\frac{3}{5}\frac{[2+(D/L)]\delta}{[1+(D/L)+\delta\{5+3(D/L)\}]}\right\} \tag{19}$$

Note that the right-hand side of Eq. (19) is independent of $h_{below}$ and $\lambda$, in agreement with the observations. The correction term is negative because $f'$ decreases as $\theta$ increases from the decrease in the size of the lever arm and from the angular acceleration. Thus the time average should be smaller than the leading order ($\theta = 0$) expression. For the three different chains used the correction terms are no more than 19% of the leading order term. This small size demonstrates consistency within the perturbative approach.

The form of the overall factor in Eq. (19) can be readily understood. The factor $[1+\delta]$ in the denominator is approximately proportional to the rotational inertia of the link about the end in contact with the floor. The factor $[1+(D/L)-\delta]$ in the numerator is there because, when $\delta = 1+(D/L)$, the "center of percussion" of the link is at the end of the link where the force $f$ is applied. Center of percussion is a term commonly used when describing baseball bats, tennis racquets or swords; when a force is applied at the center of percussion the pivot point of the link does not move, so $f' = 0$.

The values of $h_{above}/h_{below}$ predicted by Eq. (19) from the shape of the links are given in Table 2. The measured and predicted values in Table 2 for the two ball-chains agree within the statistical uncertainties. The bar-chain has a more uniform mass distribution than the ball-chains, and so is predicted to have a larger $h_{above}/h_{below}$, in qualitative agreement with measurements. However for the bar chain the discrepancy between the predicted and measured values is large.

## IV. DISCUSSION

### A. Quantitative success of the model for $h_{above}/h_{below}$.

The model proposed here for calculating $h_{above}/h_{below}$ has no free parameters. However, while there were no adjustable parameters there were many choices made as to the structure of the model. Here are discussed some of those choices and their implications.

The prediction for $h_{above}/h_{below}$ is sensitive to the how a link is defined. For example, neglecting the thickness of the links leads to drastically different results. The choice made here, shown in Fig. 3, is the most natural one for a ball chain. But this choice is much more problematic for a bar chain. For example, when the bar chain is at rest in the container there is almost certainly some bending between any two neighboring bars. Thus the rotation of one link consisting of two such bars to a straight, vertical orientation will not be a purely planar process (as it is for two balls) but will involve rotation about at least two different axes. In addition, for a bar chain the lever arm length for the force by the container becomes harder to determine because this force probably acts along the length of the bar at rest in the container and not just on the end. Thus it is understandable that the model used here to describe the ball chain does not work as well for the bar chain.

The model for the chain speed discussed in Section III.A suggests that some type of drag force is present. Such a force would slightly lower the predicted ratios of $h_{above}/h_{below}$. In contrast, adding the mass of the connecting rods between the balls/bars to the calculation of $\delta$ in Eqs. (14) and (15) would decrease the value of $\delta$, which would slightly increase the prediction for $h_{above}/h_{below}$.

The prediction of Eq. (19) is that $h_{above}/h_{below}$ is a constant. This disagrees with the observations in Ref. 20 that $h_{above}/h_{below}$ decreases at large $h_{below}$. Their observations can be explained if an additional force such as a drag force becomes relevant at large $h_{below}$. As the analysis in Section III.A shows, simultaneous measurement of the steady-state chain speed and $h_{above}/h_{below}$ at large $h_{below}$ would determine if a drag force is present.

Given its simplicity, and the lack of free parameters, the current model is quite successful.

### B. Additional physical effects.

There are interesting physical effects that are unrelated to the prediction of $h_{above}/h_{below}$. Foremost of these is the speed of waves on the chain. Waves will be generated at both ends of the chain fountain, where it leaves the container and where it strikes the floor. The speed of these waves is given by the usual expression for waves on a string, $\sqrt{T/\lambda}$, where $T$ is the tension. For the chain fountain the tension is the largest at the top where $T \approx \lambda V^2$. Substituting this into the formula for the wave speed gives that the speed of waves is equal to the chain speed at the top of the chain. Below the top the tension is less so waves propagate more slowly than the chain moves. This means that waves generated where the chain strikes the floor cannot propagate up the chain. Waves generated in the container that travel in the same direction as the chain quickly move over the top and disappear at the floor. Waves generated in the container and propagating opposite to the chain's motion initially move more slowly than the chain and so are carried up to the top of the chain where they get stuck. This explains why it is common to observe at the top of the chain relatively long lasting distortions from a smooth curve.

### C. Using the chain fountain for teaching.

There are plenty of opportunities for additional projects involving the chain fountain. As already discussed, simultaneous measurements of steady-state heights and speeds at large $h_{below}$ would be very interesting[20]. Another possibility would be to put a force sensor under the container to directly observe $F'$, the average reaction force between the chain and the container. Another possible project would be to videotape the chain fountain and study the waves propagating along the chain. This would be a good example for discussions of subcritical and supercritical flows. If a high speed video camera is available, another possible project is to observe the interactions between the chain and the container. This would allow measurement of the critical angle, which is a new observable predicted by the model.

Another possible project would be to observe the chain fountain for a wider variety of chains. The observations reported here were primarily confined to ball type chains because this is what other investigators have used. The theoretical model suggests that, in principle, many different types of chains can produce a chain fountain. However ball type chains have an advantage in that the bending angle at each junction is restricted and so these chains tend to tangle less in the container, making a chain fountain easier to observe.

When doing a classroom demonstration of the chain fountain, I usually cover the basic calculations in Sect. III.A and only the calculation of the leading order term in Eq. (19). This takes about 15 minutes and gives a numerical value to compare with measurements made in the demonstration. The large-ball chain is used for these demonstrations because it tangles less than the small-ball chain. In contrast, for a purely conceptual presentation I use the bar chain and videos of the bullet-block experiment. The bar chain is better here because it rises the highest above the container and because the bars are visually more similar to the blocks of wood in the bullet-block experiment.

The chain fountain can be used to discuss topics in continuous mass systems such as the momentum flux, speed of waves, and subcritical and supercritical flow. For discrete mass systems it can be used to discuss topics such as the relation between forces and torques, rotation and translation, reaction forces, the center of percussion, and much more. In general, the chain fountain provides a plethora of teaching opportunities.

### V. Summary

In this paper the momentum principle, a form of Newton's laws developed for continuous mass systems, has been used to describe the chain fountain. Using this it is straightforward to deduce that the chain rises up above the container because of kicks given to the chain links by the container as they are rotated away from being horizontal and at rest in the container. The system does not conserve mechanical energy because of link-link interactions just above the container bottom. The simultaneous measurements of chain speed and $h_{above}/h_{below}$ presented here suggest that the effects of any downward force on the chain where it impacts the floor can be neglected.

Calculating $h_{above}/h_{below}$ requires a model of how the chain links interact with the bottom of the container. The model developed here extends that of previous authors[18] by including the thickness of the links and also by taking into account the time dependent motion of the link in contact with the container floor. The model gives a quantitative prediction for $h_{above}/h_{below}$, Eq. (19), that has no free parameters and no ambiguity as to what constitutes a link. The model shows that a link will lift off from the container after it has rotated by only a small angle, and this critical angle is calculated, Eq. (18). The model's predictions are quite successful for ball chains but are not particularly accurate for the bar chain. This is likely because the model does not include the effects of bending at junctions inside the link, which will be much more important for the longer links made up of bars.

**Acknowledgements**

I would like to thank Riley Smith and Tyler Spilhaus for assistance with the experiments.

## Tables

Table 1. Physical parameters of the three chains. $\lambda$ is the linear mass density. For the ball chain $D$ and $L$ are the dimensions of the chain links as defined in Fig. 3. For the bar chain $L$ is defined to be the distance between the outside hemisphere in a two bar link, while $S$ is the length of the cylinder of diameter $D$ between two hemispheres in a single bar. The quantity $\delta$ is a dimensionless measure of a link's rotational inertia as defined in Eqs. (13).

|  | $\lambda$ (g/m) | $D$ (cm) | $L$ (cm) | $S$ (cm) | $\delta$ |
|---|---|---|---|---|---|
| Small ball chain | 16.5 | 0.30 | 0.44 | 0 | 1.34 |
| Large ball chain | 34.0 | 0.44 | 0.63 | 0 | 1.33 |
| Bar chain | 22.9 | 0.30 | 1.02 | 0.30 | 0.65 |

Table 2. Measured and predicted observables for the chain fountain. The measured values are the averages of the dimensionless ratios plotted in Fig. 4. All uncertainties are statistical only. The predicted values of $h_{above}/h_{below}$, $V/V_{free-fall}$ and $\theta_c$ are calculated using Eqs. (19), (7) and (18), respectively.

|  | Measured | | Predicted | | |
|---|---|---|---|---|---|
|  | $h_{above}/h_{below}$ | $V/V_{free-fall}$ | $h_{above}/h_{below}$ | $V/V_{free-fall}$ | $\theta_c$ (degrees) |
| Small ball | 0.12±0.01 | 0.70±0.03 | 0.13 | 0.748±0.003 | 16 |
| Large ball | 0.12±0.01 | 0.69±0.03 | 0.13 | 0.748±0.003 | 16 |
| Bar | 0.14±0.01 | 0.71±0.04 | 0.32 | 0.755±0.003 | 27 |

## Figure Captions

1) Photograph of a chain fountain. Here the rise height is 0.60 m, the fall height is 3.6 m and the chain is the bar type.
2) Photograph showing the three chains used in this experiment.
3) Sketch showing one link of the chain being lifted from the container, with the forces acting on it and how the lengths are labeled.
4) Measured, dimensionless ratios as a function fall height, $h_{below}$, are plotted for three different chains. The upper plot shows the ratio of the rise height to the fall height ($h_{above}/h_{below}$) and the lower plot shows the ratio of the steady-state chain speed to the free-fall speed ($V/V_{free-fall}$). Note that (1) the dimensionless ratios are approximately constant, independent of fall height and (2) the bar chain rises slightly higher than the other chains.
5) Sketch of forces acting on the chain fountain (color online).

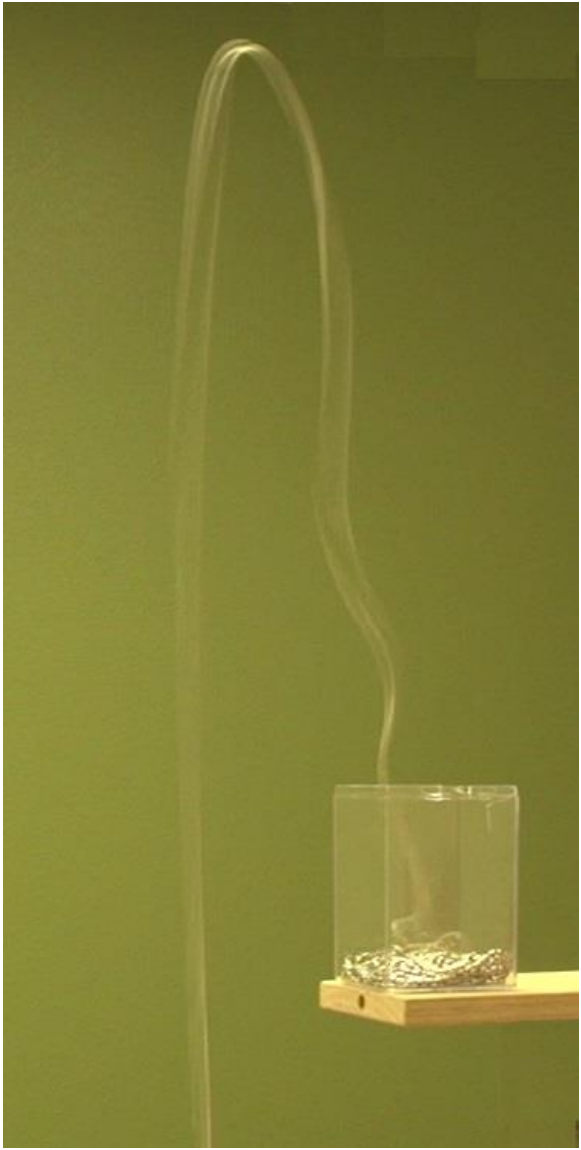

Fig. 1.

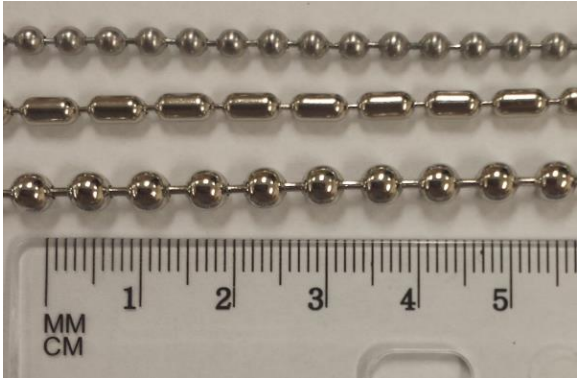

Fig. 2

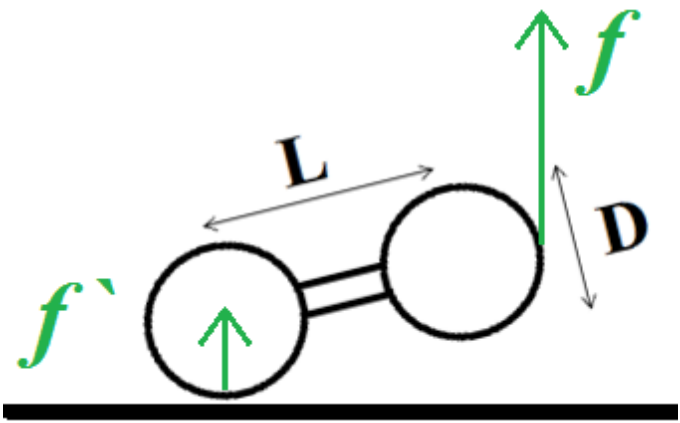

Fig. 3

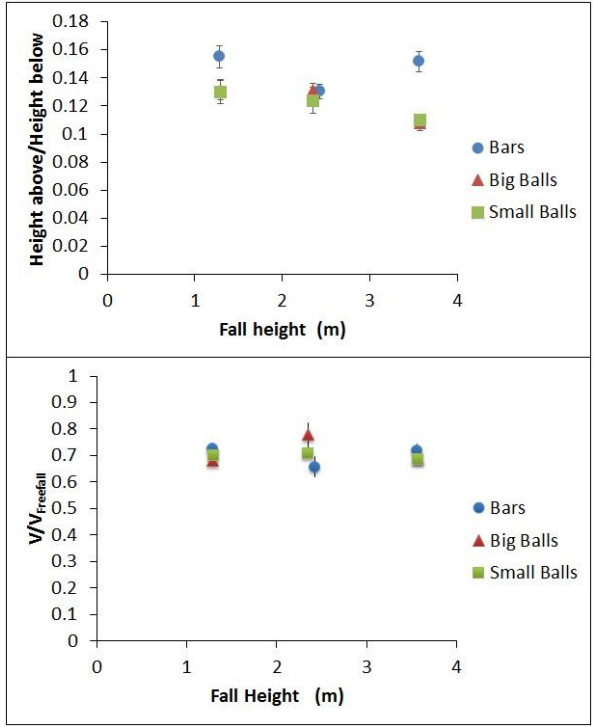

Fig. 4

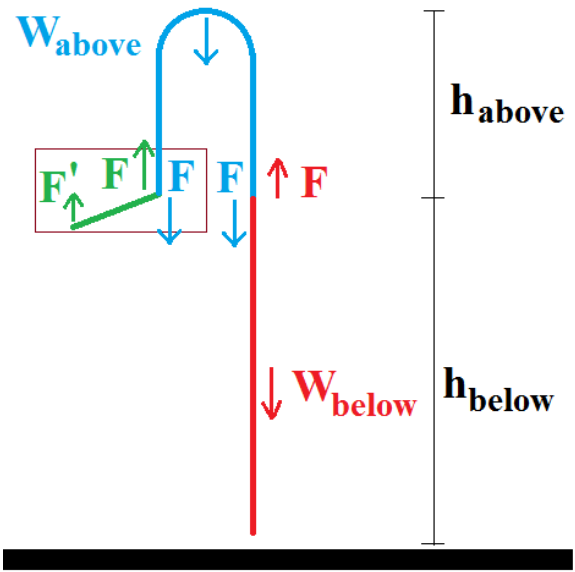

Fig. 5